Title Page

# Unveiling Early Warning Signals of Systemic Risks in Banks: A Recurrence Network-Based Approach

Shijia Song
202131250022@mail.bnu.edu.cn
School of Systems Science, Beijing Normal University, Beijing 100875, China

Handong Li
lhd@bnu.edu.cn
School of Systems Science, Beijing Normal University, Beijing 100875, China

Correspondence should be addressed to Handong Li
E-mail: lhd@bnu.edu.cn
Tel. 8610-58807064
Tax. 8610-58807876

# Unveiling Early Warning Signals of Systemic Risks in Banks: A Recurrence Network-Based Approach

Abstract: Bank crisis is challenging to define but can be manifested through bank contagion. This study presents a comprehensive framework grounded in nonlinear time series analysis to identify potential early warning signals (EWS) for impending phase transitions in bank systems, with the goal of anticipating severe bank crisis. In contrast to traditional analyses of exposure networks using low-frequency data, we argue that studying the dynamic relationships among bank stocks using high-frequency data offers a more insightful perspective on changes in the banking system. We construct multiple recurrence networks (MRNs) based on multidimensional returns of listed banks' stocks in China, aiming to monitor the nonlinear dynamics of the system through the corresponding indicators and topological structures. Empirical findings indicate that key indicators of MRNs, specifically the average mutual information, provide valuable insights into periods of extreme volatility of bank system. This paper contributes to the ongoing discourse on early warning signals for bank instability, highlighting the applicability of predicting systemic risks in the context of banking networks.

## 1. Introduction

Banking crises represent costly and frequently recurring events. According to the definition provided by Basel III, common risk types prevalent in the current banking market include credit risk, market risk, and operational risk. While these risk types pertain to individual-level bank risks, when individual banks are unable to bear such risks, they can transmit them to other banks through direct or indirect business connections, thereby triggering crises in other banks. This cascade effect exposes numerous banks to simultaneous shocks, giving rise to what is known as systemic risk in the banking sector (Hasman, 2013). Over the past few decades, more than two-thirds of countries in the International Monetary Fund (IMF) have experienced such financial crises in various forms. The banking crises reveal the fact that not only is the banking sector unable to transfer problems originating from a single economic sector, but it is also a major driver of financial crises itself.

Current research on such banking system crises primarily focuses on discussions concerning the causes of banking crises (Boyd & Runkle, 1993; Demsetz et al., 1997; Diamond & Dybvig, 1983; Lehuede et al., 2012; Nakamura & Roszbach, 2018), studies on the mechanisms of banking crisis contagion based on network theory, and research on measuring banking system risk. Since the pioneering work of Leitner (2005), which highlighted contagion risk as a driving force behind network design, viewing the banking system as a network composed of interconnected institutions and influenced by endogenous and exogenous fluctuations has become commonplace (Haldane & May, 2011; May et al., 2008). Babus (2016), Ladley (2013), Allen et al. (2012), Sachs (2011), and others discuss the impact of network structure on risk contagion, while Nier et al. (2007) analyze the resilience of different banking market structures to systemic risk. Subsequently, mainstream research has depicted the response of banking networks to systemic risk through simulations of external shocks (Acemoglu et al., 2015; Li & Wang, 2021; Roukny et al., 2018; Tasca et al., 2017) analyze the contribution of individual banks to systemic risk and the differences in the contribution of indirect and direct risk contagion between banks.

Regarding the measurement of banking system risk, the most common methods currently rely on tail risk models based on stock market returns and network-based structural metrics. Specific systemic risk measures under tail risk models include CoVaR method, MES method, SRISK method, nonlinear Granger test, and others (Acharya et al., 2017; Adrian & Brunnermeier, 2016; Brownlees & Engle, 2017). These models can provide timely and accurate insights into the actual risk conditions of the banking market, but their results are contingent on the efficiency of the securities market. On the other hand, the measurement of systemic risk based on networks depends on the structural analysis of exposure networks and statistical analysis of correlation networks (Battiston et al., 2016). The former requires real trading data or confidential regulatory data, which is generally difficult to obtain, and most analyses are one-off, making it challenging to analyze the actual evolution of the network (Hasman, 2013). The latter usually involves the analysis of publicly available market data, establishing relevant networks for banks and revealing risks through changes in indicators (Anand et al., 2015; Cont et al., 2013).

Despite the increasing richness of research on banking crises and systemic risk, the majority of studies have been focused on investigating the causes of risk contagion, simulating contagion pathways, conducting stress tests, and measuring systemic risk. Few scholars have undertaken theoretical and empirical research with the specific purpose of predicting banking system crises. However, monitoring the risk evolution of banking systems and uncovering potential Early Warning Signals (EWS) for crises hold significant practical significance. Researchers such as Guttal et al. (2016) and Diks et al. (2019) find that an increase in the volatility of relevant financial indicators could provide warning signals for financial crises. Nevertheless, Diks et al. (2019) point out that these indicators may not be reliable EWS due to the complexity of financial markets, which typically involve multiple variables and parameters. Squartini et al. (2013) confirm that sudden crises lead to significant structural changes in the interbank network's topology, and these structural changes serve as precursors to approaching crises. Saracco et al. (2016) and Joseph et al. (2014) adopt similar methods in their studies, but these research efforts generally described observed warning signals qualitatively rather than quantitatively, limiting their practical applications. Quax et al.

(2013) introduce the concept of Information Dissipation Length (IDL) as a leading indicator of global instability in dynamic systems. Gatfaoui and Peretti (2019) analyze information propagation patterns in dynamic space-time networks, where nodes are connected through short-term causal relationships. They observe flickering information propagation before reaching a critical point, but they find that the occurrence of financial crises is far from a deterministic critical point, making it difficult to provide precise early warning signals for real banking crises.

As evident, the task of early warning for banking system crises or systemic risk faces considerable challenges, while offering ample research opportunities. In this backdrop, many scholars have sought to move beyond using purely financial econometric models or simple network models to study such issues. Rather, they aim to explore methods and theories from the emerging field of econophysics to investigate the risk evolution of these complex systems. Hasselman (2022) argues that existing mainstream methods based on time series analysis often employ crude data handling by eliminating non-stationarity, heterogeneity, and long memory, considering them as interfering factors (Molenaar, 2004; Ramachandran, 1979). This approach inadvertently weakens the underlying complexity of data, potentially hindering the precise capturing of system state transitions. Thus, Hasselman (2022) proposes the use of multiplex recurrence networks (MRNs) to quantify the dynamics of complex systems while allowing for the presence of non-stationarity and heterogeneity. He confirms that structural metrics of such networks can act as potential EWS to reveal system phase transitions. Prior research by Lacasa et al. (2015) has already demonstrated that peak indicators from MRNs have distinct roles in stable and unstable periods in financial systems. Similarly, Eroglu et al. (2018) employs critical MRN indicators to characterize the evolution of dynamic systems in the vegetation context. However, their focus was not on predicting system crises but on studying the correspondence between these indicators' changes and event occurrences.

Considering the strong complexity of financial systems, this paper emphasizes the considerable potential of the MRN method in early warning for banking system risks and seeks to promote the application of this technique to risk monitoring and prediction in banking systems. In selecting data, following the viewpoints in the IMF working paper by Chen and Svirydzenka (2021), macro indicators like the credit-to-GDP ratio do not appear to be the best predictors for banking system crises. Instead, changes in stock prices and credit gaps have shown promise in providing early warnings for banking crises in advanced economies, and stock prices and real estate prices have proven useful for early warning in emerging market banking crises. Moreover, as mentioned earlier, traditional measurements of banking systemic risk are mostly based on debt relationships and common risk exposure among banks, and those constructed based on stock prices essentially measure the collective fluctuations of listed bank stocks. The former often involves long time scales, sometimes using years for representation, and has certain limitations in risk monitoring and management. The latter, however, can be based on high-frequency data to express banking system fluctuations or risks on a daily scale. Therefore, this paper considers the use of high-frequency stock price data for listed banks, aiming to establish a multiplex recurrence network for the banking system. By employing this non-linear tool, the objective is to capture the internal non-linear complex relationships within the banking system. Through the dynamic evolution of the network, the paper aims to monitor and provide early warnings for systemic risk at smaller time scales, facilitating risk monitoring and prediction for banking institutions.

The remaining sections are arranged as follows: Chapter 2 provides a detailed introduction to the modeling method of MRN and the quantification tools for measuring the extreme volatility of bank system. Chapter 3 employs Monte Carlo simulation to determine the optimal parameters for the MRN indicators in effectively indicating the banking system's state. Chapter 4 conducts an empirical analysis using stock data from China's listed banks to demonstrate the superiority of the proposed approach in providing better early warnings for banking system crises compared to mainstream methods. Chapter 5 summarizes.

## 2. Method

### 2.1 Construction of multiplex recurrence network

Multiplex recurrence networks (MRNs) are derived from single recurrence networks (RNs), and these RNs serve as network-theory-based extensions of recurrence plots (RPs).Eckmann et al. (1987) introduced the concept of RP, considering it to be a fundamental property of all dynamic systems. RP is a two-dimensional visualization tool developed to represent this property. Over the last two decades, RP has evolved into a nonlinear method for describing complex dynamics (Marwan et al., 2007).

RP visually represents the recurring states of a dynamic system in its $m$-dimensional phase space. To determine whether phase space vectors $\vec{x_i}(i = 1, \ldots, N, \vec{x_i} \in \mathcal{R}^m)$ are close, a pairwise measurement is performed on their distances, denoted by $R_{i,j}$ as follows:

$$R_{i,j} = \Theta\left(\varepsilon - d(\vec{x_i}, \vec{x_j})\right), \quad (1)$$

where $\Theta(\cdot)$ is the Heaviside function, $\varepsilon$ is a threshold for $d(\vec{x_i}, \vec{x_j})$, and $d(\vec{x_i}, \vec{x_j})$ represents the distance between the vectors (Marwan, 2007). The distance measure, $d(\vec{x_i}, \vec{x_j})$, can be determined using various methods such as spatial distance, string metrics, or local rank order (Bandt et al., 2008; Marwan et al., 2007). Typically, spatial distance is represented by the Euclidean norm, i.e., $d(\vec{x_i}, \vec{x_j}) = \|\vec{x_i} - \vec{x_j}\|$. The binary recurrence matrix, denoted as $R$, has elements $R_{i,j}$, where $R_{i,j}$ is set to 1 if the distance $\|\vec{x_i} - \vec{x_j}\|$ is less than $\varepsilon$, and 0 otherwise. The phase space trajectory can be

reconstructed from the time series $\{u_i\}_{i=1}^{N}$ using time-delay embedding (Packard et al., 1980):

$$\vec{x_i} = (u_i, u_{i+\tau}, \ldots, u_{i+\tau(m-1)}), \quad (2)$$

where $m$ denotes the embedding dimension and $\tau$ represents the time delay, and their appropriate values can be chosen based on Abarbanel (1996). Empirically, $\varepsilon$ is often chosen to yield a recurrence rate of 0.05 or as 0.05 times the standard deviation of the original data (Eroglu et al., 2018; Marwan et al., 2007).

The RP exhibits a diagonal line representing the recurrence of each point with itself, and if spatial distance is used for recurrence, the RP is symmetric. The diagonal and vertical lines in the RP reveal small-scale features. Zbilut and Webber (1992) introduced Recurrence Quantification Analysis (RQA), which quantitatively describes RP's line structures. It defines measures such as diagonal line segment properties, recurrence rate, average length of diagonal structures, entropy, and more (Webber, & Marwan, 2015), providing insights into the analyzed system's nonlinear dynamics. However, RQA's focus on identifying recurrence patterns in the time series might not fully capture the complexity of the dependence between variables. Simultaneously, single-variable time series often fail to offer a comprehensive representation of high-dimensional complex systems. To facilitate a more profound structural analysis of such complexity and to extend single-dimensional time series analysis to the multidimensional domain, the utilization of multiple recurrent networks has emerged.

An $M$-layer multiplex recurrence network is created by combining $M$ recurrence networks. Given an $M$-dimensional time series $\{s(t)\}_{t=1}^{N}$, where $s(t) = (s_1(t), s_2(t), \cdots, s_M(t)) \in R^M$, we can build $M$ distinct recurrence networks, each comprising $N$ nodes representing different time points. These networks constitute separate layers of the multiplex recurrence network, and they are interconnected based on the nonlinear relationships between corresponding time nodes in each network.

To construct the multiplex recurrence network, we follow Eq.(1) to generate the recurrence networks for the $M$ components of $s(t)$ and obtain the corresponding adjacency matrices $A^{[k]} = a_{ij}^{[k]}$, where $a_{ij}^{[k]} = 1$ (i.e., $R_{ij}^{[k]} = 1$) if the $i^{th}$ and the $j^{th}$ vertices in the $k^{th}$ layer are connected, and 0 otherwise. The giant adjacency matrix describing the multiplex network can be expressed as:

$$\mathcal{A} = \begin{bmatrix} A^{[1]} & E_N & \ldots & E_N \\ E_N & A^{[2]} & \ddots & E_N \\ \vdots & \ddots & \ddots & E_N \\ E_N & \ldots & E_N & A^{[M]} \end{bmatrix}, \quad (3)$$

where $E_N$ is the identity matrix of size $N$. In practice, the construction of the multiplex recurrence network does not rely on the giant $NM \times NM$ matrix but rather on representing the $M$-layer recurrence networks as interconnected vertices. As a result, the MRN becomes a weighted network of size $M \times M$. This projection of multilayer networks into single-layer networks has been demonstrated to be a superior and efficient approach compared to methods requiring symbolic processing of time series

(Eroglu et al., 2018; Hasselman, 2022; Lacasa et al., 2015).

The interlayer mutual information (Lacasa et al., 2015) is a widely used measure to capture the nonlinear dependence between different layers. It quantifies the correlations between the degrees of the same node at layers $\alpha$ and $\beta$, where $\alpha, \beta = 1, \cdots, M;\ \alpha \neq \beta$. The interlayer mutual information $I_{\alpha,\beta}$ is defined as:

$$I_{\alpha,\beta} = \sum_{k^{[\alpha]}} \sum_{k^{[\beta]}} P(k^{[\alpha]}, k^{[\beta]}) \ln \frac{P(k^{[\alpha]}, k^{[\beta]})}{P(k^{[\alpha]}) P(k^{[\beta]})}, \tag{4}$$

where $P(k^{[\alpha]})$ and $P(k^{[\beta]})$ are the degree distributions of recurrence networks at layers $\alpha$ and $\beta$, respectively, and $P(k^{[\alpha]}, k^{[\beta]})$ is the joint distribution of vertices with degree $k^{[\alpha]}$ at layer $\alpha$ and degree $k^{[\beta]}$ at layer $\beta$.

## 2.2 The indicators of MRN

The indicators of MRN serve as a quantification tool to characterize the information shared across all layers of the high-dimensional system. The averaged mutual information ($I$) is defined as:

$$I = \langle I_{\alpha,\beta} \rangle = \frac{1}{M} \sum_{\alpha > \beta} I_{\alpha,\beta}, \tag{5}$$

which represents the typical information flow between the multivariate time series and is regarded as an important structural measure of MRNs (Eroglu et al., 2018; Lacasa et al., 2015).

Another common measure that captures the overall coherence of the original time series in the MRN is the Average Edge Overlap ($\omega$). We calculate the proportion of edges shared between any two vertices in each layer-pair and then compute the average value across all layer-pairs. Thus, the Average Edge Overlap $\omega$ can be defined as:

$$\omega_{\alpha,\beta} = \frac{\sum_i \sum_{j>i} \left(a_{ij}^{[\alpha]} + a_{ij}^{[\beta]}\right)}{M \sum_i \sum_{j>i} \left(1 - \delta_{0, a_{ij}^{[\alpha]} + a_{ij}^{[\beta]}}\right)} \tag{6}$$

$$\omega = \langle \omega_{\alpha,\beta} \rangle = \frac{1}{M} \sum_{\alpha > \beta} \omega_{\alpha,\beta}. \tag{7}$$

This measure also represents the global coherence between different layers of the MRN and, together with the averaged mutual information, forms the most significant quantitative indicators of MRN used to detect phase transitions in multidimensional time series systems. A sudden increase in these two indicators signifies an abrupt rise in the similarity between different dimensional variables in the system. Such an occurrence may indicate the emergence or impending emergence of specific system behaviors, thus providing a theoretical basis for its potential use as an early warning signals (EWS)( Lacasa et al.,2015; Eroglu et al.,2018; Hasselman, 2022).However, existing studies have yielded inconsistent conclusions regarding methods to quantify the mutations of the indicators and mainly rely on visual observation, which makes it challenging to identify peaks in indicator sequences of larger lengths. In this study, we calculate the differences between adjacent indices in the indicator sequences of length $T$ and determine its $95^{th}$ percentile as $\Delta\xi$. We define peaks in the indicator sequence as $\xi_p$ if and only if the following conditions are simultaneously satisfied for all $p$ in $T$: $(\xi_p - \xi_{p-1})(\xi_p - \xi_{p+1}) > 0$, $\xi_p - \xi_{p-1} > \Delta\xi$ and $\xi_p - \xi_{p+1} > \Delta\xi$.

## 2.3 Quantification of banking crises

Despite numerous studies that have explained the definition and causes of bank crises and simulated their contagion pathways, a clear quantitative tool for identifying the bank system's crisis remains elusive when only publicly available market data is accessible. Even tools like CoVaR and SRISK, which focus on describing the magnitude of systemic risk, do not have clear standards to determine the level of risk that can be considered as systemic risk. In this study, we consider a representative portfolio of listed banks in the market as a concrete representation of the banking system in that market. The volatility of the portfolio's returns is regarded as an indicator of the system's risk exposure. Specifically, we treat jumps in the portfolio returns as bank system crises and consider using the classical BNS jump test to identify the day on which the bank crisis occurs.

Barndorff-Nielsen et al.(2004) introduced the bipower variation measure of daily returns, $BV_T$, as a reliable estimate of the population variance of the price process. Consequently, the contribution of jumps to the population variance can be estimated as the difference between realized volatility and bipower variation, denoted as $RV_T - BV_T$. The corresponding statistic for jumps is denoted as $Z_{BNS,T}$, and it follows a standard normal distribution $Z_{BNS,T} \xrightarrow{D} N(0,1)$. The expression for $Z_{BNS,T}$ is defined as follows:

$$Z_{BNS,T} = \frac{1 - \frac{BV_T}{RV_T}}{\sqrt{\left(\left(\frac{\pi}{2}\right)^2 + \pi - 5\right)M^{-1}\max\left(1, \frac{TP_T}{BV_T^2}\right)}}, \tag{8}$$

where $RV_T = \sum_{i=1}^{M} r_{T_i}^2$, $BV_T = \frac{\pi}{2}\left(\frac{M}{M-1}\right)\sum_{i=2}^{M}|r_{T_i}||r_{T_{i-1}}|$, $TP_T = \mu_{4/3}^{-3}\left(\frac{M^2}{M-2}\right)\sum_{i=3}^{M}|r_{T_{i-2}}|^{\frac{4}{3}}|r_{T_{i-1}}|^{\frac{4}{3}}|r_{T_i}|^{\frac{4}{3}}$, $\mu_{4/3} = 2^{\frac{2}{3}}\Gamma(7/6)\Gamma(1/2)^{-1}$. Here, $i$ represents the $i^{th}$ high-frequency moment within day $T$, and there are $M$ high-frequency returns within a day. It is important to note that the returns mentioned in this article are assumed to be logarithmic returns, defined as the difference between logarithmic prices, i.e., $r_{T_i} = logP_{T_i} - logP_{T_{i-1}}$, where $P$ represents the price of the bank portfolio. The significance level of the BNS test can greatly influence the number of detected jumps,

i.e., the quantity of bank crises. In the next section, we will use simulation experiments to determine at which confidence level the indicators of the multiplex recurrence network perform best in indicating extreme fluctuations in the banking system.

## 3. Simulation experiment

In this section, we will conduct a simple Monte Carlo simulation based on the price mechanism model. The objective is to observe the peaks of MRN indicators correspond most accurately to banking system crises at which level of risk. The core of this simulation involves generating high-frequency multidimensional price sequences with a time step of 1 minute. The window length is set to 240 minutes, equivalent to the trading duration of a single trading day in China's stock market. Within each window, the multidimensional price series is used to construct the MRN and calculate the corresponding indicators, thereby generating daily-level indicator sequences. In each experiment, we vary the frequency of co-jumps occurring across the multiple price sequences, which is also considered as the occurrence frequency of system crises or the risk level. By employing the method of identifying peaks introduced in Section 2.2, we can determine the peaks in the indicator sequences and evaluate the performance of the indicator peaks in indicating actual co-jumps using sensitivity and specificity measures (Altman & Bland, 1994). Subsequently, following the principles of the ROC curve (Brown & Davis, 2006), which involve minimizing $(1 - specificity)^2 + (sensitivity - 1)^2$ in a specific experiment, we can consider the corresponding co-jump frequency as a relatively optimal risk level. In the empirical part, we will evaluate the effectiveness of the method proposed in this paper for predicting banking system crises under this risk level.

Assuming that asset prices follow a general jump-diffusion process, in existing classical models of the price process, the diffusion process is typically assumed to be a functional form of Brownian motion. In this case, we assume the existence of an exogenous jump component that affects prices, and its arrivals follow a compound Poisson process. In discrete form, we consider the price process consisting of two parts: one part is the independent incremental process of prices, and the other part is the compound Poisson process representing co-jumps. The interaction of these two parts leads to changes in prices over time. The stochastic differential equation form of the asset price model is as follows:

$$p_t = p_0 + dp_t + \sum_{k=0}^{N_t} J_k \tag{9}$$

$dp(t)$ can be represented by a geometric Brownian motion process as the incremental part of prices: $dp_t = \mu dt + \sigma dB(t)$, where $B(t)$ represents the standard Brownian motion, $\mu$ and $\sigma$ are the annual yield and volatility of prices, assumed to be constant, and $dt$ is the unit of time relative to one year. $\sum_{k=0}^{N_t} J_k$ is the compound

Poisson process, and we assume $J_k$ is an i.i.d. uniform distribution variable representing the magnitude of jumps. The arrivals of jumps follow a Poisson process with an arrival rate of $\lambda$, and all jumps are differentiated in positive and negative directions. The definition of returns is consistent with the previous text.

According to the above price dynamics equation, we use Monte Carlo simulation to generate 5 sequences of asset prices. We assume that the unit step of the series is 1 minute and the number of trading days is 250 (approximately a year). Since there are 240 trading minutes on each day in China's stock market, we generate 600001 simulated prices and thus obtain 60000 1-minute returns. Table 1 lists the parameters involved in the simulation and their values.

Table 1. Parameter setting of Monte Carlo simulation

| Parameter | Value |
| --- | --- |
| Annual return | $\sim U(0.1, 0.2)$ |
| Annual volatility | $\sim U(0.1, 0.3)$ |
| Initial price | $\sim U(500, 1000)$ |
| Probability of co-jumps | [0.001, 0.05] with increasing step of 0.001 of each experiment for both positive and negative jumps |
| Size $J$ of the positive cojump | $\sim U(10, 15)$ |
| Size $J$ of the negative cojump | $\sim U(15, 20)$ |

The ranges of all parameters are selected based on the actual situation of the Chinese stock market and are denoted in RMB. Building on the findings of Campbell and Hentschel (1992), Bollerslev et al.(2006), which demonstrated stronger volatility for negative returns than for positive returns, we assign the average magnitude of negative jumps to be greater than the average magnitude of positive jumps.

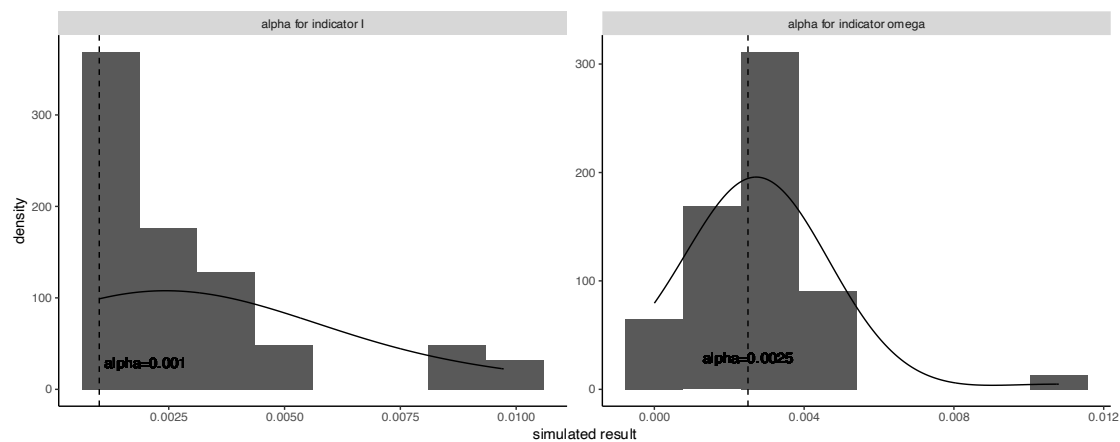

Figure 1. The distributions of optimal risk level for two MRN indicators based on simulation experiments.

Due to the high computational complexity of the simulation, this study conducted the above operations only 50 times. At this point, the results' distribution has reached a relatively stable state. Figure 1 presents the distribution of optimal risk levels obtained from different simulation experiments. The graph reveals that for indicator $I$, its indication of system co-jumps is most effective at a frequency of approximately 0.001, while for indicator $\omega$, its indication of system co-jumps is most effective at a frequency of approximately 0.0025. Consequently, in the subsequent empirical analysis, we will perform BNS tests on the banking system at confidence levels of $\alpha = 0.001$ and $\alpha = 0.0025$, respectively. We will then observe the early warning effects of these two indicators in identifying the so-called banking crises.

## 4. Empirical analysis

### 4.1 Data

In this study, the listed banks in the China's stock market are considered as potential empirical objects. To ensure data length, we take the year 2012 as the starting point and consider all banks listed on the Shanghai and Shenzhen stock exchanges from 2012 onwards as representatives of China's banking system. The 1-minute closing price sequences of each listed bank from January 4, 2012, to October 30, 2020, are selected as observational values for this system in different dimensions. After basic data cleaning, the historical data covers 2,005 trading days. The market capitalization of each listed bank in 2020 is normalized and used as weight values, forming a weighted combination of each bank's stock prices as the representation of the banking system's prices. Table 2 presents the empirical objects of this study along with their respective stock codes.

Table 2. Empirical objects and the corresponding stock codes

| Bank | Code | Bank | Code |
| --- | --- | --- | --- |
| Industrial And Commercial Bank Of China Co., Ltd. | sh601398 | China Everbright Bank Company Limited Co., Ltd | sh601818 |
| China Construction Bank Co., Ltd. | sh601939 | Industrial Bank Co.,Ltd. | sh601166 |
| Agricultural Bank Of China Co., Ltd. | sh601988 | Shanghai Pudong Development Bank Co.,Ltd. | sh600000 |
| Bank Of China Co., Ltd. | sh601288 | Ping An Bank Co., Ltd. | sz000001 |
| Bank of Communications Co.,Ltd. | sh601328 | Hua Xia Bank Co.,Ltd. | sh600015 |
| China Merchants Bank Co., Ltd. | sh600036 | Bank Of Beijing Co.,Ltd. | sh601169 |
| China Minsheng Banking Co., Ltd. | sh600016 | Bank Of Nanjing Co.,Ltd. | sh601009 |
| China Citic Bank Co.,Ltd. | sh601998 | Bank Of Ningbo Co.,Ltd. | sz002142 |

## 4.2 Performance of MRN indicators

Similarly, using 240 minutes as the unit window length, we establish MRNs for each window length and obtain the daily-level indicator sequences $\{I_t\}_{t=1}^{2005}$ and $\{\omega_t\}_{t=1}^{2005}$, from which we identify the peaks of the indicator sequences. When applying the BNS test to the banking system crises, considering the possible impact of market microstructure noise and following the sampling principle of "often but not too often" (Andersen, Bollerslev, Diebold, & Ebens, 2001; Andersen, Bollerslev, Diebold, & Labys, 2001), we choose a sampling frequency of 5 minutes for high-frequency returns. At $\alpha = 0.001$, we detect 12 extreme fluctuations in the banking system, and at $\alpha = 0.0025$, we detect 27 occurrences of banking crises. Figures 2 and 3 display the early warning effects of the peaks of indicators $I$ and $\omega$ on banking system crises. We consider the moments when the indicator peaks occur and the subsequent 9 days as the risk intervals for potential banking crises, as indicated by the color blocks in Figure 2.

From Figure 2, it can be observed that 9 out of the 12 banking crises are included in the predicted unstable periods, with 3 of them precisely coinciding with the peaks. This indicates that the average mutual information has a good early warning effect on banking crises. On the other hand, Figure 3 shows that the average edge overlap successfully predicts only 7 out of the 27 banking system crises, showing a less satisfactory performance as a warning signal. The underlying reason may be that the average edge overlap primarily measures linear correlations between multiple layers of recurrence networks, while average mutual information captures more non-linear relationships between different dimensions in complex systems, making it more sensitive to emerging behaviors. Therefore, in the subsequent discussion, we will pay more attention to the average mutual information.

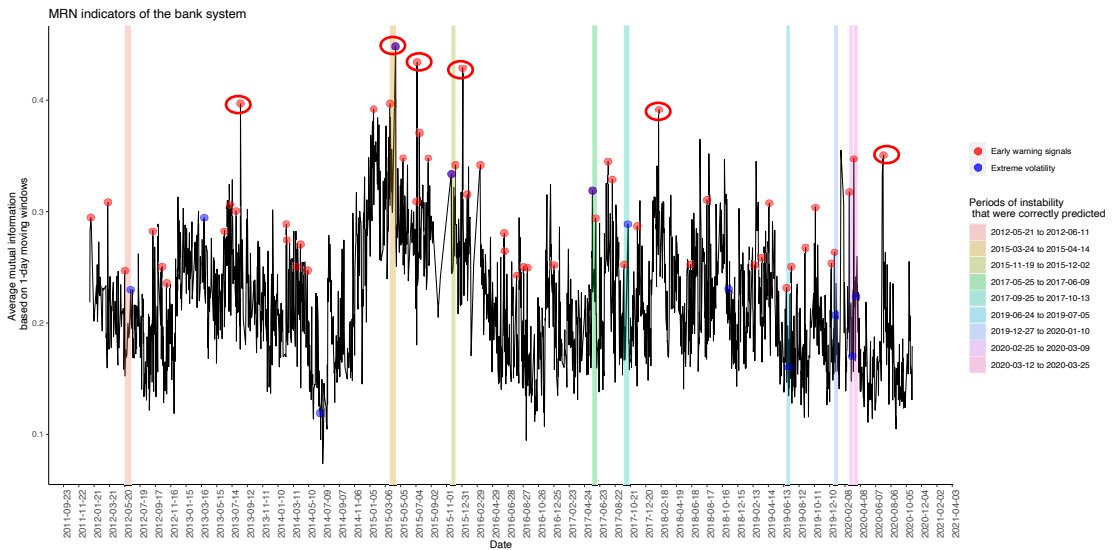

Figure 2. EWSs and unstable periods of China's bank system indicated by peaks of average mutual information. The blue dots represent jumps in the bank portfolio detected by BNS test at a confidence level of 0.001. The red dots represent the identified peaks in the average mutual information sequence. The 9 days following the appearance of the red dots, as well as the day marked by the red dots, are considered potentially unstable periods where extreme fluctuations of the banking system may occur.

In order to further assess the potential of the average mutual information as an early warning signal for banking system crises, we employ a benchmark model using the same dataset and compare the findings. Although there is a relative scarcity of research on early warning signals for banking system crises, there is no universally accepted benchmark in this field. However, Guttal et al.(2016) propose utilizing abrupt changes in variance, mean power spectrum, and autocorrelation function at lag-1 of returns in the period preceding a financial crash as an early warning signal for such events. This proposition has gained widespread recognition among scholars. Hence, we adopt this method as our benchmark, and its core idea can be summarized as follows: First, we identify the highest price points before the crisis and apply a Gaussian kernel smoothing function to remove the long-term trend within a certain window (set as 500 days) before these highest price points to obtain residuals for subsequent analysis. We set the length of the rolling window and calculate the corresponding sequences of lag-1 autocorrelation, variance, and low-frequency power spectrum based on the rolling window. Next, we select a segment from the indicator sequence and estimate its Kendall's $\tau$ rank correlation coefficient with the sequence $\{1,2,\ldots,lkw\}$ to determine whether the indicator exhibited an upward or downward trend within the year before the crisis, where $lkw$ refers to the length of the selected segment, and $lkend$ is defined as the distance from the endpoint of this segment to the highest price point before the crash. A positive (negative) Kendall's $\tau$ indicates an upward (downward) trend in the indicator. As this process involves four parameters, detailed sensitivity analysis is required to ensure the robustness of the conclusions. Thus, for each indicator, the sample distribution of Kendall's $\tau$ is obtained, which indicates whether the choice of parameters directly affects the effectiveness of the indicator trend. Specifically, if the histogram shows clear peaks close to 1 (or -1), it suggests that the trend of the corresponding indicator is robust to parameter changes and exhibits a strong upward (or downward) trend.

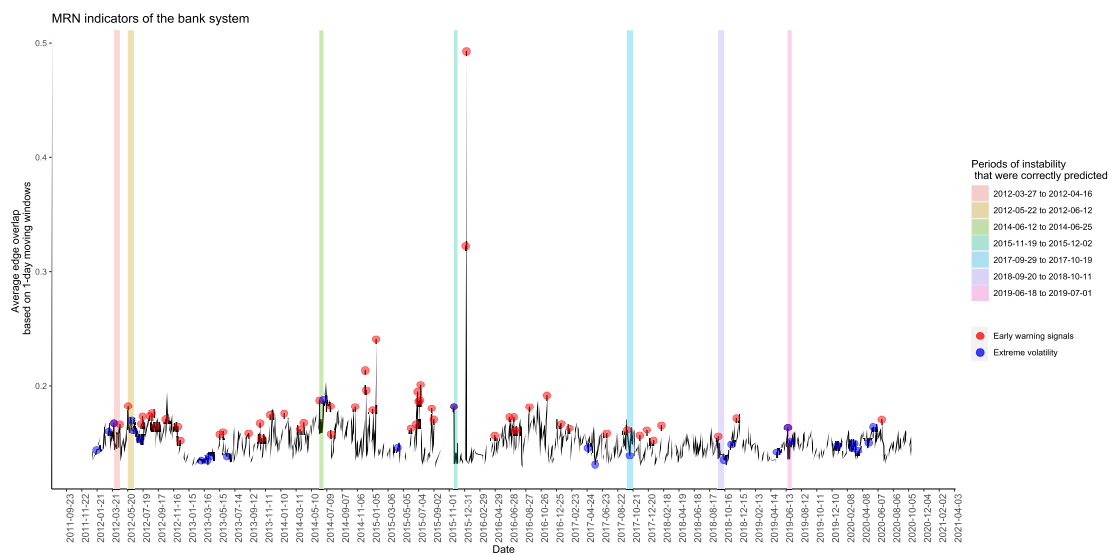

Figure 3. EWSs and unstable periods of China's bank system indicated by peaks of average edge overlap. The blue dots represent jumps in the bank portfolio detected by BNS test at a confidence level of 0.001. The red dots represent

the identified peaks in the average mutual information sequence. The 9 days following the appearance of the red dots, as well as the day marked by the red dots, are considered potentially unstable periods where extreme fluctuations of the banking system may occur.

In addition to sensitivity analysis, we also pay attention to whether the estimated Kendall's $\tau$ before the crisis is significant compared to more distant history. To do this, we calculate all indicators and their Kendall's $\tau$ with the same parameter values in a historical window of 1000 days before the crisis. The p-value determining the significance of Kendall's $\tau$ is defined as the proportion of Kendall's $\tau$ in this sample that are greater than or equal to the Kendall's $\tau$ of the data within one year before the stock market crash. If this p-value is less than 0.1, it indicates that the trend of the indicators before the banking crisis is significant at the 10% confidence level and has early warning signal (EWS) effect. Otherwise, it cannot be considered an effective EWS indicator.

During the benchmark testing, we consider the 12 extreme fluctuation points identified by the BNS as known banking system crises and then perform the aforementioned procedures. Figure 4 illustrates the results of three crises as examples. The graph demonstrates that prior to the bank crisis on November 19, 2015, the Kendall's $\tau$ distribution of mean power spectrum and autocorrelation function at lag-1 shows a trend approaching 1, indicating that these indicators pass the sensitivity test and exhibit an upward trend before the occurrence of extreme risks in the banking system. Similarly, before the crisis on May 25, 2017, the Kendall's $\tau$ distribution of mean power spectrum and autocorrelation function at lag-1 shows a trend approaching -1, indicating that these indicators also pass the sensitivity test and demonstrate a downward trend before the occurrence of extreme risks. However, other events and indicators do not pass the sensitivity test.

However, upon further analysis, it is evident that the changes in mean power spectrum and autocorrelation function at lag-1 within this narrow time window, when compared to the relatively extensive historical data, do not meet the criteria for statistical significance. As a result, they cannot be regarded as significant early warning indicators.

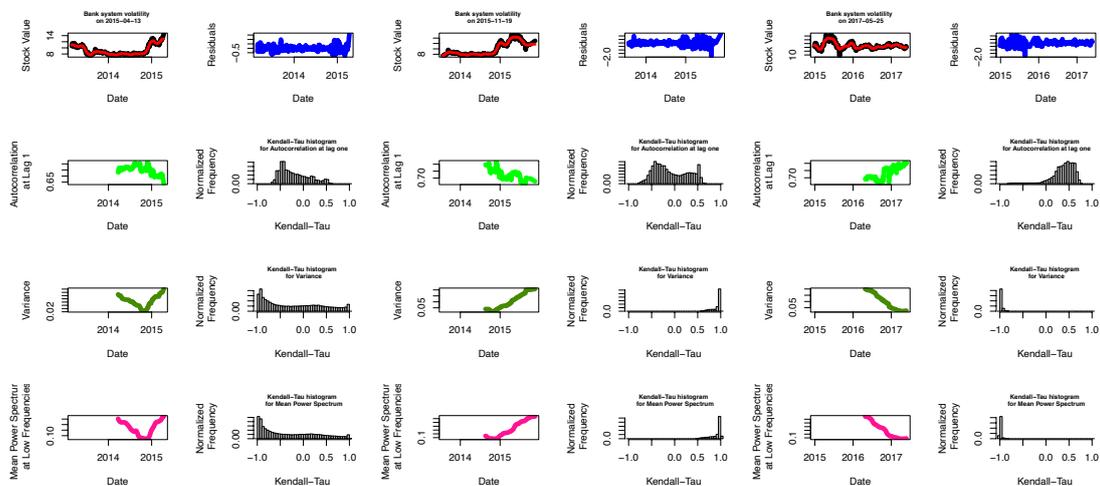

Figure 4. Early warning signals of benchmark for three major bank crises of 2015 and 2017. Each of the two columns indicates the result corresponding to one of the crisis. The first row presents the price trend of bank portfolio for the 4 years (1000 days) prior to the crisis and the corresponding detrended residuals.

After analyzing all 12 known bank crises, we present the benchmark results in Table 2 to determine whether it can effectively provide early warning signals for these crises.

Table 2. Ratio of benchmark indicators passing sensitivity and significance tests

|  | Sensitivity test | Significance test |
|---|---|---|
| variance | 0/12 | 1/12 |
| mean power spectrum | 2/12 | 6/12 |
| autocorrelation function at lag-1 | 2/12 | 5/12 |

The results in the table indicate that the benchmark indicators do not effectively provide early warning signals for extreme fluctuations in the banking system. In contrast, the average mutual information based on the MRN offers more reliable and accurate early warning signals under the same conditions. In practical market scenarios, this indicator holds significant implications for supervising the banking system and preventing banking system crises.

## 4.3 Topological analysis of multiple recurrence networks

The multiple recurrence network not only provides a potential indicator that can serve as an early warning signal for banking system crises but also its own topological structure is worth further in-depth analysis. Based on the moments corresponding to the peaks of the average mutual information identified in Figure 2, we trace back to the corresponding maximum spanning tree structure of the MRN and discover a fundamental pattern: the maximum spanning tree of the MRN corresponding to peaks generally lacks a dominant core node, indicating a disordered state. In contrast, the maximum spanning tree of the MRN corresponding to non-peaks moments usually exhibits a central node, and this central node remains relatively stable over a certain period of time. The left side of Figure 5 shows the maximum spanning tree structures of the most evident six peaks (marked with red circles) from Figure 2, while the right side shows the topological structures of randomly selected six non-peaks periods corresponding to the MRNs. This discovery facilitates our transition from the macroscopic analysis to the examination of individual banks, enabling us to investigate which banks hold dominant positions in the banking system (such as Industrial Bank, ICBC, China Construction Bank, and China Merchants Bank) and the subtle changes in the banking network's topological structure before crises occur.

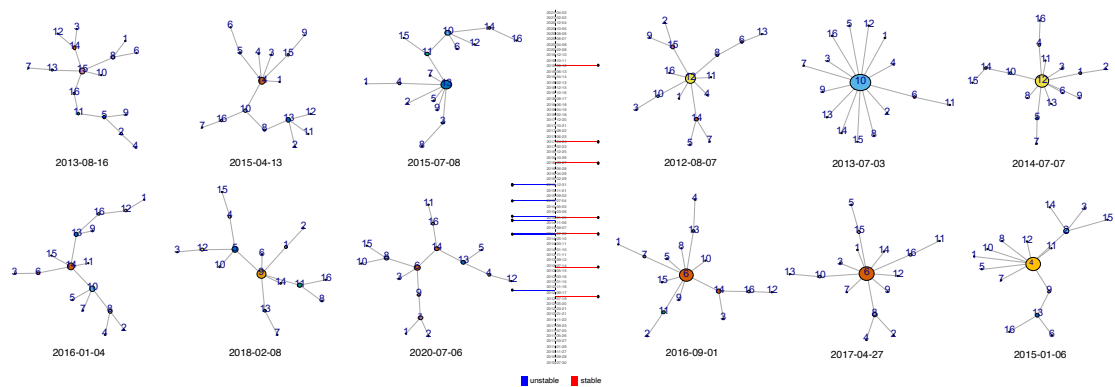

Figure 5. The Maximum Spanning Trees of the corresponding MRNs of China's banking system. The MSTs on the left of the time horizon belong to the so-called period of systemic instability, while the MSTs on the right belong to the period of systemic stability. Each constituent stock is assigned the same color in all the networks, while the size of a node is proportional to its degree.

In addition, during the six periods corresponding to the peaks circled in Figure 2, the entire market experienced extreme fluctuations, which are as follows: a)August 16, 2013: A trading error incident occurred at China Everbright Bank, known as the "Everbright Wulong Incident." b) April 13, 2015: The Shanghai Composite Index fell by 7.7%, marking the largest single-day decline since the global financial crisis in 2007, breaking the 4000-point.c)July 8, 2015: The Shanghai Composite Index experienced a brief plunge of over 8%, and the Shenzhen Component Index also declined by more than 8%, setting another record for the largest single-day decline, intensifying market panic.d)January 4, 2016: China's stock market triggered the circuit breaker mechanism.e) February 8, 2018: The Shanghai Composite Index fell by 5.05%, marking the largest single-day decline since January 2016. The Shenzhen Component Index also dropped by 5.31%. f) July 6, 2020: The Chinese stock market experienced a significant decline, with the Shanghai Composite Index falling by nearly 4%, and the ChiNext Index dropping by over 5%. This finding suggests that the topological structure of the multiple recurrence network can also reveal extreme fluctuations in the entire market to some extent, not just limited to the banking system's fluctuations.

However, it also indicates that there were hardly any extreme fluctuations in the banking system near the most prominent peaks of the average mutual information. This finding suggests that China's banking system is relatively robust. Even during periods of external market turbulence, the banking system shows resilience and is less susceptible to extreme fluctuations. This observation further implies that banking system risk or volatility is not solely manifested through market price fluctuations; rather, it is more likely influenced by the emergence of nonlinear correlations or nonlinear similarity behavior among banks.

# 5. Conclusion

This study has successfully investigated the potential application of multiplex recurrence-based analytical methods in monitoring early warning signals (EWSs) of the extreme volatility of bank system in financial markets. Our empirical results indicate that the peaks of basic indicators of MRNs, i.e., the average mutual information, are effective in providing EWSs for bank crises and show a superior strength compared with the commonly used benchmark model.

The main contribution of this paper lies in providing a comprehensive framework for monitoring EWSs before phase transitions of bank system, incorporating a quantitative method for detecting the peaks of MRN-based indicators and a jump test for identifying the realized bank crisis. This framework retains key characteristics of time series data, such as non-stationarity and long memory, without requiring dimensionality reduction, preserving the maximum complexity of the data, which maybe the core reason for its higher accuracy in capturing the evolution of bank system compared to existing mainstream EWS indicators. Moreover, the MRN topology could unveiling the period of instability and stability of bank system in some extent. And it also reveals the fact that China's banking system is relatively stable compared to volatile external markets.

However, it is essential to acknowledge some limitations in our study. The use of publicly available market data may present constraints in accurately capturing all aspects of systemic risk of banks, and the effectiveness of our proposed approach could be further enhanced by incorporating more comprehensive datasets and accounting for potential interdependencies between different financial institutions.In future research, expanding the scope of data sources and applying our method to diverse financial markets could offer valuable comparative insights and a broader perspective on analysis of bank crisis.

Overall, this study lays the groundwork for continued research into systemic risk and bank crises, contributing to the broader goal of enhancing financial stability and resilience in the face of market uncertainties. By combining innovative methodologies and empirical evidence, we hope to provide valuable support for policymakers, financial institutions, and investors in making informed decisions to safeguard bank systems.

Reference

Abarbanel, H. D. I. (1996). Choosing the Dimension of Reconstructed Phase Space. In

> H. D. I. Abarbanel (Ed.), *Analysis of Observed Chaotic Data* (pp. 39–67).

> Springer. https://doi.org/10.1007/978-1-4612-0763-4_4

Acemoglu, D., Ozdaglar, A., & Tahbaz-Salehi, A. (2015). Systemic Risk and Stability

in Financial Networks. *American Economic Review*, *105*(2), 564–608. https://doi.org/10.1257/aer.20130456

Acharya, V. V., Pedersen, L. H., Philippon, T., & Richardson, M. (2017). Measuring Systemic Risk. *Review of Financial Studies*, *30*(1), 2–47. https://doi.org/10.1093/rfs/hhw088

Adrian, T., & Brunnermeier, M. K. (2016). CoVaR. *American Economic Review*, *106*(7), 1705–1741. https://doi.org/10.1257/aer.20120555

Allen, F., Babus, A., & Carletti, E. (2012). Asset commonality, debt maturity and systemic risk. *Journal of Financial Economics*, *104*(3), 519–534. https://doi.org/10.1016/j.jfineco.2011.07.003

Altman, D. G., & Bland, J. M. (1994). Diagnostic tests. 1: Sensitivity and specificity. *BMJ : British Medical Journal*, *308*(6943), 1552.

Anand, K., Craig, B., & von Peter, G. (2015). Filling in the blanks: Network structure and interbank contagion. *Quantitative Finance*, *15*(4), 625–636. https://doi.org/10.1080/14697688.2014.968195

Andersen, T. G., Bollerslev, T., Diebold, F. X., & Ebens, H. (2001). The distribution of realized stock return volatility. *Journal of Financial Economics*, *61*(1), 43–76. https://doi.org/10.1016/S0304-405X(01)00055-1

Andersen, T. G., Bollerslev, T., Diebold, F. X., & Labys, P. (2001). The Distribution of Realized Exchange Rate Volatility. *Journal of the American Statistical Association*, *96*(453), 42–55. https://doi.org/10.1198/016214501750332965

Babus, A. (2016). The formation of financial networks. *The RAND Journal of*

*Economics*, *47*(2), 239–272.

Bandt, C., Groth, A., Marwan, N., Romano, M. C., Thiel, M., Rosenblum, M., & Kurths, J. (2008). Analysis of Bivariate Coupling by Means of Recurrence. *Mathematical Methods in Signal Processing and Digital Image Analysis*, 153–182. https://doi.org/10.1007/978-3-540-75632-3_5

Barndorff-Nielsen, O. E. (2004). Power and Bipower Variation with Stochastic Volatility and Jumps. *Journal of Financial Econometrics*, *2*(1), 1–37. https://doi.org/10.1093/jjfinec/nbh001

Battiston, S., Farmer, J. D., Flache, A., Garlaschelli, D., Haldane, A. G., Heesterbeek, H., Hommes, C., Jaeger, C., May, R., & Scheffer, M. (2016). Complexity theory and financial regulation. *Science*, *351*(6275), 818–819. https://doi.org/10.1126/science.aad0299

Bollerslev, T., Litvinova, J., & Tauchen, G. (2006). Leverage and Volatility Feedback Effects in High-Frequency Data. *Journal of Financial Econometrics*, *4*(3), 353–384. https://doi.org/10.1093/jjfinec/nbj014

Boyd, J. H., & Runkle, D. E. (1993). Size and performance of banking firms: Testing the predictions of theory. *Journal of Monetary Economics*, *31*(1), 47–67. https://doi.org/10.1016/0304-3932(93)90016-9

Brown, C. D., & Davis, H. T. (2006). Receiver operating characteristics curves and related decision measures: A tutorial. *Chemometrics and Intelligent Laboratory Systems*, *80*(1), 24–38. https://doi.org/10.1016/j.chemolab.2005.05.004

Brownlees, C., & Engle, R. F. (2017). SRISK: A Conditional Capital Shortfall Measure

of Systemic Risk. *The Review of Financial Studies*, *30*(1), 48–79.

Campbell, J. Y., & Hentschel, L. (1992). No news is good news: An asymmetric model of changing volatility in stock returns. *Journal of Financial Economics*, *31*(3), 281–318. https://doi.org/10.1016/0304-405X(92)90037-X

Chen, S., & Svirydzenka, K. (2021). *Financial Cycles – Early Warning Indicators of Banking Crises?* (SSRN Scholarly Paper 3970194). https://doi.org/10.2139/ssrn.3970194

Cont, R., Moussa, A., & Santos, E. B. (2013). Network Structure and Systemic Risk in Banking Systems. In J.-P. Fouque & J. A. Langsam (Eds.), *Handbook on Systemic Risk* (pp. 327–368). Cambridge University Press. https://doi.org/10.1017/CBO9781139151184.018

Demsetz, R. S., Saidenberg, M. R., & Strahan, P. E. (1997). *Agency Problems and Risk Taking at Banks* (SSRN Scholarly Paper 943507). https://doi.org/10.2139/ssrn.943507

Diamond, D. W., & Dybvig, P. H. (1983). Bank Runs, Deposit Insurance, and Liquidity. *Journal of Political Economy*, *91*(3), 401–419.

Diks, C., Hommes, C., & Wang, J. (2019). Critical slowing down as an early warning signal for financial crises? *Empirical Economics*, *57*(4), 1201–1228. https://doi.org/10.1007/s00181-018-1527-3

Eckmann, J.-P., Kamphorst, S. O., & Ruelle, D. (1987). Recurrence Plots of Dynamical Systems. *Europhysics Letters*, *4*(9), 973. https://doi.org/10.1209/0295-5075/4/9/004

Eroglu, D., Marwan, N., Stebich, M., & Kurths, J. (2018). Multiplex recurrence networks. *Physical Review E*, *97*(1), 012312. https://doi.org/10.1103/PhysRevE.97.012312

Gatfaoui, H., & de Peretti, P. (2019). Flickering in Information Spreading Precedes Critical Transitions in Financial Markets. *Scientific Reports*, *9*(1), Article 1. https://doi.org/10.1038/s41598-019-42223-9

Guttal, V., Raghavendra, S., Goel, N., & Hoarau, Q. (2016). Lack of Critical Slowing Down Suggests that Financial Meltdowns Are Not Critical Transitions, yet Rising Variability Could Signal Systemic Risk. *PLOS ONE*, *11*(1), e0144198. https://doi.org/10.1371/journal.pone.0144198

Haldane, A. G., & May, R. M. (2011). Systemic risk in banking ecosystems. *Nature*, *469*(7330), Article 7330. https://doi.org/10.1038/nature09659

Hasman, A. (2013). A Critical Review of Contagion Risk in Banking. *Journal of Economic Surveys*, *27*(5), 978–995. https://doi.org/10.1111/j.1467-6419.2012.00739.x

Hasselman, F. (2022). Early Warning Signals in Phase Space: Geometric Resilience Loss Indicators From Multiplex Cumulative Recurrence Networks. *Frontiers in Physiology*, *13*. https://www.frontiersin.org/articles/10.3389/fphys.2022.859127

Joseph, A. C., Joseph, S. E., & Chen, G. (2014). Cross-border Portfolio Investment Networks and Indicators for Financial Crises. *Scientific Reports*, *4*(1), Article 1. https://doi.org/10.1038/srep03991

Lacasa, L., Nicosia, V., & Latora, V. (2015). Network structure of multivariate time series. *Scientific Reports*, *5*(1), Article 1. https://doi.org/10.1038/srep15508

Ladley, D. (2013). Contagion and risk-sharing on the inter-bank market. *Journal of Economic Dynamics and Control*, *37*(7), 1384–1400. https://doi.org/10.1016/j.jedc.2013.03.009

Lehuede, H. J., Kirkpatrick, G., & Teichmann, D. (2012). *Corporate Governance Lessons from the Financial Crisis* (SSRN Scholarly Paper 2393978). https://doi.org/10.2139/ssrn.2393978

Leitner, Y. (2005). Financial Networks: Contagion, Commitment, and Private Sector Bailouts. *The Journal of Finance*, *60*(6), 2925–2953. https://doi.org/10.1111/j.1540-6261.2005.00821.x

Li, S., & Wang, C. (2021). Network structure, portfolio diversification and systemic risk. *Journal of Management Science and Engineering*, *6*(2), 235–245. https://doi.org/10.1016/j.jmse.2021.06.006

Marwan, N., Carmen Romano, M., Thiel, M., & Kurths, J. (2007). Recurrence plots for the analysis of complex systems. *Physics Reports*, *438*(5), 237–329. https://doi.org/10.1016/j.physrep.2006.11.001

May, R. M., Levin, S. A., & Sugihara, G. (2008). Ecology for bankers. *Nature*, *451*(7181), Article 7181. https://doi.org/10.1038/451893a

Molenaar, P. C. M. (2004). A Manifesto on Psychology as Idiographic Science: Bringing the Person Back Into Scientific Psychology, This Time Forever. *Measurement: Interdisciplinary Research and Perspectives*, *2*(4), 201–218.

https://doi.org/10.1207/s15366359mea0204_1

Nakamura, L. I., & Roszbach, K. (2018). Credit ratings, private information, and bank monitoring ability. *Journal of Financial Intermediation*, *36*, 58–73. https://doi.org/10.1016/j.jfi.2017.11.001

Nier, E., Yang, J., Yorulmazer, T., & Alentorn, A. (2007). Network models and financial stability. *Journal of Economic Dynamics and Control*, *31*(6), 2033–2060. https://doi.org/10.1016/j.jedc.2007.01.014

Packard, N. H., Crutchfield, J. P., Farmer, J. D., & Shaw, R. S. (1980). Geometry from a Time Series. *Physical Review Letters*, *45*(9), 712–716. https://doi.org/10.1103/PhysRevLett.45.712

Quax, R., Kandhai, D., & Sloot, P. M. A. (2013). Information dissipation as an early-warning signal for the Lehman Brothers collapse in financial time series. *Scientific Reports*, *3*(1), Article 1. https://doi.org/10.1038/srep01898

Ramachandran, B. (1979). On the "Strong Memorylessness Property" of the Exponential and Geometric Probability Laws. *Sankhyā: The Indian Journal of Statistics, Series A (1961-2002)*, *41*(3/4), 244–251.

Roukny, T., Battiston, S., & Stiglitz, J. E. (2018). Interconnectedness as a source of uncertainty in systemic risk. *Journal of Financial Stability*, *35*, 93–106. https://doi.org/10.1016/j.jfs.2016.12.003

Sachs, A. (2011). *Completeness, Interconnectedness and Distribution of Interbank Exposures—A Parameterized Analysis of the Stability of Financial Networks* (SSRN Scholarly Paper 1997182). https://doi.org/10.2139/ssrn.1997182

Saracco, F., Di Clemente, R., Gabrielli, A., & Squartini, T. (2016). Detecting early signs of the 2007–2008 crisis in the world trade. *Scientific Reports*, *6*(1), Article 1. https://doi.org/10.1038/srep30286

Squartini, T., van Lelyveld, I., & Garlaschelli, D. (2013). Early-warning signals of topological collapse in interbank networks. *Scientific Reports*, *3*(1), Article 1. https://doi.org/10.1038/srep03357

Tasca, P., Battiston, S., & Deghi, A. (2017). Portfolio diversification and systemic risk in interbank networks. *Journal of Economic Dynamics and Control*, *82*, 96–124. https://doi.org/10.1016/j.jedc.2017.01.013

Webber, C. L., & Marwan, N. (Eds.). (2015). *Recurrence Quantification Analysis: Theory and Best Practices*. Springer International Publishing. https://doi.org/10.1007/978-3-319-07155-8

Zbilut, J. P., & Webber, C. L. (1992). Embeddings and delays as derived from quantification of recurrence plots. *Physics Letters A*, *171*(3), 199–203. https://doi.org/10.1016/0375-9601(92)90426-M